\documentclass[%
reprint,  
superscriptaddress,
 amsmath,amssymb,
 aps,
prb,
]{revtex4-2}
\usepackage{graphicx}
\usepackage{dcolumn}
\usepackage{bm}
\usepackage[usenames,dvipsnames,sgvnames,table]{xcolor} 
\usepackage[T1]{fontenc}
\usepackage{placeins}
\usepackage{hyperref}

\begin{document}

\preprint{APS/123-QED}

\title{
Electric and Magnetic Field-dependent Tunneling between Coupled Nanowires
}

\author{Shashank Anand}
\affiliation{Department of Physics and Astronomy, Purdue University, West Lafayette, IN 47907, USA}
\affiliation{Purdue Quantum Science and Engineering Institute, West Lafayette, IN 47907, USA}

\author{Ranjani Ramachandran}
\affiliation{Department of Physics and Astronomy, University of Pittsburgh, Pittsburgh, Pennsylvania 15260, USA}
\affiliation{Pittsburgh Quantum Institute, Pittsburgh, Pennsylvania 15260, USA}

\author{Kitae Eom}
\affiliation{Department of Materials Science and Engineering, University of Wisconsin–Madison, Madison, Wisconsin 53706, USA}

\author{Kyoungjun Lee}
\affiliation{Department of Materials Science and Engineering, University of Wisconsin–Madison, Madison, Wisconsin 53706, USA}

\author{Dengyu Yang}
\affiliation{Department of Physics and Astronomy, University of Pittsburgh, Pittsburgh, Pennsylvania 15260, USA}
\affiliation{Pittsburgh Quantum Institute, Pittsburgh, Pennsylvania 15260, USA}

\author{Muqing Yu}
\affiliation{Department of Physics and Astronomy, University of Pittsburgh, Pittsburgh, Pennsylvania 15260, USA}
\affiliation{Pittsburgh Quantum Institute, Pittsburgh, Pennsylvania 15260, USA}

\author{Sayanwita Biswas}
\affiliation{Department of Physics and Astronomy, University of Pittsburgh, Pittsburgh, Pennsylvania 15260, USA}
\affiliation{Pittsburgh Quantum Institute, Pittsburgh, Pennsylvania 15260, USA}

\author{Aditi Nethwewala}
\affiliation{Department of Physics and Astronomy, University of Pittsburgh, Pittsburgh, Pennsylvania 15260, USA}
\affiliation{Pittsburgh Quantum Institute, Pittsburgh, Pennsylvania 15260, USA}

\author{Chang-Beom Eom}
\affiliation{Department of Materials Science and Engineering, University of Wisconsin–Madison, Madison, Wisconsin 53706, USA}

\author{Erica W. Carlson}
\email{ewcarlson@purdue.edu}
\affiliation{Department of Physics and Astronomy, Purdue University, West Lafayette, IN 47907, USA}
\affiliation{Purdue Quantum Science and Engineering Institute, West Lafayette, IN 47907, USA}

\author{Patrick Irvin}
\affiliation{Department of Physics and Astronomy, University of Pittsburgh, Pittsburgh, Pennsylvania 15260, USA}
\affiliation{Pittsburgh Quantum Institute, Pittsburgh, Pennsylvania 15260, USA}

\author{Jeremy Levy}
\email{jlevy@pitt.edu}
\affiliation{Department of Physics and Astronomy, University of Pittsburgh, Pittsburgh, Pennsylvania 15260, USA}
\affiliation{Pittsburgh Quantum Institute, Pittsburgh, Pennsylvania 15260, USA}


\date{\today}

\begin{abstract}
Coupled quasi-one-dimensional (quasi-1D) electron systems host rich emergent physics that cannot be accounted for by understanding isolated 1D electron systems alone. 
Open questions remain about 
how transport in these arrays can be manipulated by the application of external electric and magnetic fields. In this theoretical study, we consider a pair of coupled nanowires with non-interacting electrons.  We find that a metal-insulator transition is induced by an out-of-plane magnetic field and a transverse potential bias on an array of such coupled wires. We demonstrate the existence of distinct conductance features and highlight the crucial role played by the field dependence of the interwire potential barrier on transport properties. These predictions agree well with transport experiments performed on coupled nanowires sketched on an $\mathrm{LaAlO}_3/\mathrm{SrTiO}_3$ interface. Since our model makes minimal assumptions, we expect our predictions to hold for a wide class of coupled 1D systems.
\end{abstract}
\maketitle


 \section{Introduction}
  Many exotic quantum phases can arise from quasi-one-dimensional electronic systems (1DES). For instance, quantum Hall phases have been predicted to exist for arrays of quantum wires in a magnetic field \cite{kane2002,teo2014}. Extensive work mapping non-Fermi liquid phases in anisotropic materials to interacting Luttinger liquids is also gaining attention~\cite{vishwanath2001, yu2023, wang2022, wen1990}, and sliding Luttinger liquid phases and spin liquid phases of matter are also predicted to exist for coupled 1DESs~\cite{mukhopadhyay2001,mukhopadhyay2001a,meng2015,patel2016}. In addition, much work has been done on the connection between unconventional superconductivity and stripe phases\cite{emery2000,kivelson1998,carlson2002,zaanen2000}. Numerical techniques like density matrix renormalization group performed on a Hubbard 2-leg ladder  point to distinct instabilities in two-dimensional systems \cite{shen2023,noack1997,jiang2022}.

Engineered arrays of coupled nanowires offer an important testing ground for quantum simulation of such emergent phenomena. 
For example, $\mathrm{LaAlO}_3/\mathrm{SrTiO}_3$ (LAO/STO) nanowires have been shown to have remarkably long mean-free paths, and regimes of metallic and superconducting pairing behavior \cite{annadi2018,pai2018,levy2022}. The interface itself hosts a range of non-Fermi liquid properties including ferroelasticity, ferromagnetism \cite{pai2018,levy2022} and superconductivity\cite{caviglia2008,pai2018,levy2022}. The fact that these properties manifest in 
nanowires that are tunable makes them good candidates for performing analog quantum simulations of various correlated phases as well as strong contenders for the construction of nanoelectronic devices \cite{levy2022}. 
Prior work exploring the tunneling physics of coupled nanowires
largely focused on physics arising due to the energy spectrum of the wires, 
with less attention paid to the nature of the interwire potential 
landscape\cite{wolf2011,boese2001,jiang1994,jesusa.delalamo1995,tserkovnyak2002}. This assumption is valid if the potential barrier is much larger than all other energy scales in the system. However, the presence of large electric and magnetic fields can alter the interwire potential landscape which, in turn, influences the tunneling rate significantly.
For example, it has  been shown for parabolic quantum wells, crossed electric and magnetic fields gives rise to quasi-bound states~\cite{santiago2003} that form between adjacent wells, which influence tunneling current. 

In this paper, we explore the effects of a magnetic and electric field-dependent interwire potential barrier on the transverse differential conductance (henceforth referred to as ``conductance'')
of coupled nanowires and discuss the physics behind the distinctive features that arise. The following features arise in our model:
(1) There is a critical voltage above which the system starts to conduct in the transverse direction, and the value of the critical voltage depends on the applied transverse magnetic field.  
This voltage demarcates distinct metallic and insulating phases in an extended array of nanowires. The similarity between the conductance map and the density of states activated for tunneling as a function of the applied electric and magnetic fields  allows us to interpret and understand the physics of the system itself. 
(2) Characteristic peaks and valleys in the transverse conductance maps follow from a field tunable interwire potential landscape and momentum dependent shift in the guiding centers of electrons. Specifically, we identify a local minimum in the transverse conductance  $G_{\perp}(B)$ at $B=0$ and 
$V_T = 0$
as a consequence of the magnetic field dependence
of the interwire potential. 
(3) The effects of magnetic depopulation are prominent in wires with multiple occupied subbands. They give rise to sharp changes in conductance that are independent of electric field. 
(4) Increasing the potential barrier relative to the chemical potential opens up an insulating region in the conductance at low fields.


We compare our findings to  transport experiments on nanostructures sketched on the LAO/STO interface. Recent work~\cite{annadi2018,briggeman2020,cen2010,pai2018,levy2022} has demonstrated the ability to draw conducting nanostructures 
into the LAO/STO interface, down to a resolution of  $\sim 10~$nm \cite{yang2020}.
When such nanostructures are constructed in parallel and gated, it is possible to explore the physics of tunneling between nanowires separated by a potential barrier that is tuned by the application of electric and magnetic fields. We demonstrate that many important trends 
observed experimentally on coupled nanowires drawn on LAO/STO
are consistent with the findings of our model. 
Because our model makes minimal assumptions, these  generic features 
are expected to apply 
to a large class of coupled electron waveguides beyond the specific LAO/STO system to which we compare. 

\begin{figure}
     \centering
     \includegraphics[scale=1]{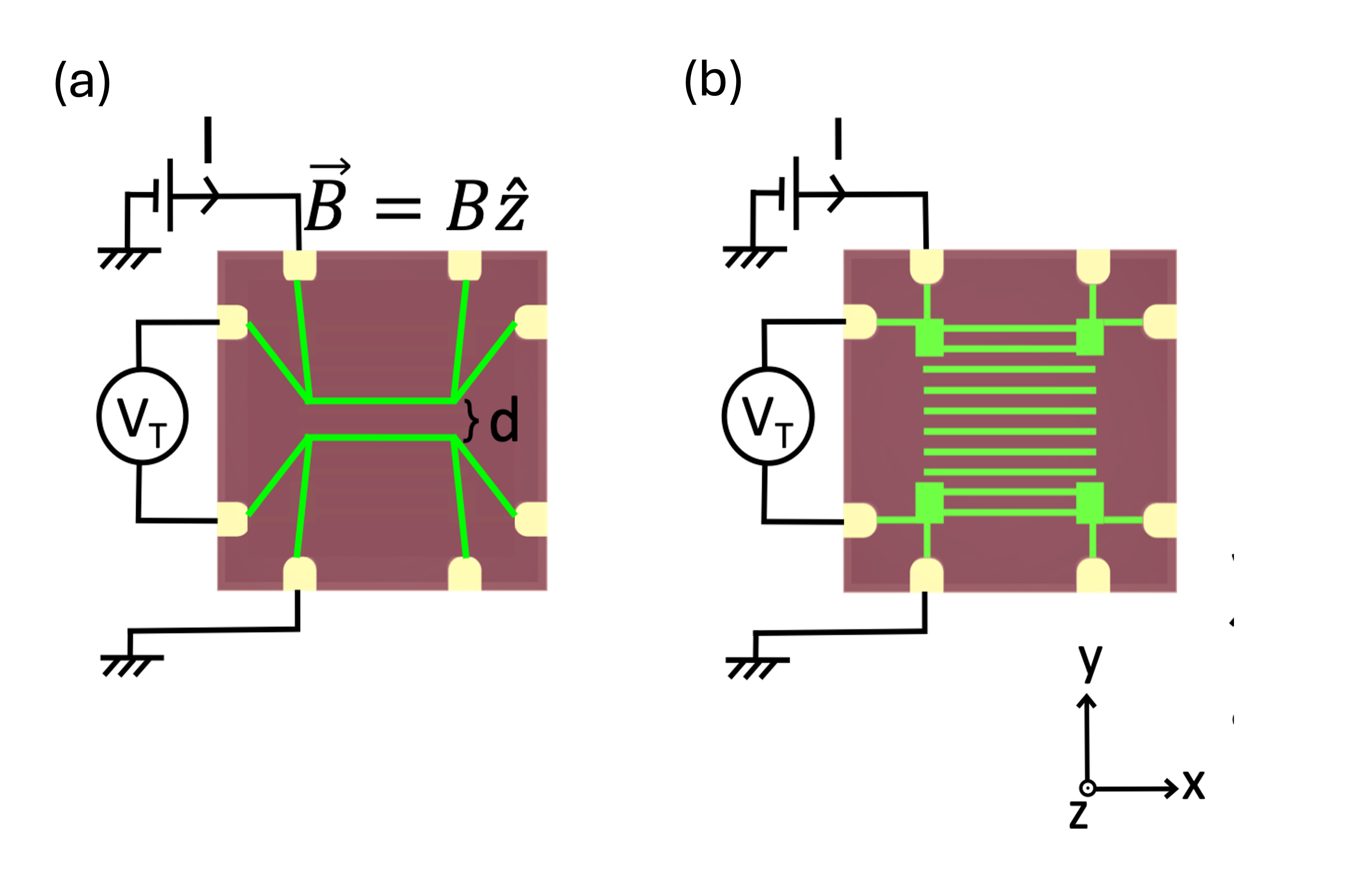}
     \caption{(a)~ Two nanowires placed in parallel a distance $d$ apart with a transverse bias $V_T$ and out of plane magnetic field $B$. (b) An array of parallel nanowires a distance $d$ apart from each other with a transverse bias $V_T$ and out of plane magnetic field $B$.}
     \label{fig:circ}
 \end{figure}
\section{Coupled electron waveguides}
We consider two identical coupled electron waveguides, each in the regime of Landauer quantized ballistic transport~\cite{annadi2018}.
We treat the direction along a given wire ($\hat{x}$) as translationally invariant while electrons
experience a confining potential  along the transverse and out of plane directions, $(\hat{y})$ and $(\hat{z})$, respectively. We assume that bands are isotropic along the $x$-$y$ plane making $m_x^*=m_y^*=m^*$.  
We treat the in-plane confinement direction $(\hat{y})$ in the harmonic approximation.
We model the confinement in the out of plane ($\hat{z}$ direction) via a half-parabola, with $V(z)=\frac{1}{2}m^*_z\omega_z^2z^2$ for $z\leq0$ and $V(z)=\infty$ for $z>0$. Thus, for $z\leq0$, the Hamiltonian for a single wire in Landau gauge is written as:
\begin{align}
    \begin{split}
      \mathcal{H}_{wire}&=\frac{1}{2m^*}(p_x-eBy)^2+\frac{p_y^2}{2m^*}+\frac{p_z^2}{2m_z^*
      }  \\& \quad+\frac{1}{2}m^*\omega_y^2y^2 +\frac{1}{2}m^*_z \omega_z^2z^2  -g\mu_BBs \label{eqn:hwire}
    \end{split}    
\end{align}
where $p_i$  is the canonical momentum operator along the $i$th axis, g is the 
Landé g-factor, $\mu_B$ is the Bohr magneton, and $s=\pm 1/2$ is the spin of the electron. The angular frequency of the harmonic potentials is $\omega_i=\hbar/m^*_il_i^2$, where $l_i$ characterizes the confinement length of the electron along the $i$th axis. 

The eigenenergies of a single wire are
\begin{align}
\begin{split}
E_{n}(B)&=(n_y+1/2)\hbar \Omega_B+((2n_z+1)+1/2)\hbar\omega_z  \\& \quad -
g\mu_B Bs+\frac{\hbar^2k_x^2\omega_y^2}{2m^*\Omega_B^2}~.  \label{eqn:en} 
\end{split}
\end{align}
The in-plane frequency is shifted by the application of an out of plane magnetic field, and is given by $\Omega_B=\sqrt{(\hbar/m^*l_y^2)^2+(eB/m^*)^2}$. The wire's transverse subbands are indexed by $n_y \in \mathbb{N}$. 
Annadi {\em et al.} have shown that this model accounts well for the observed
transport behavior of single nanowires drawn on the LAO/STO platform \cite{annadi2018}.
We assume that only the lowest z-subband is occupied, and set $n_z = 0$ for the remainder of the paper.
The physics of a single wire is discussed further in Appendix \ref{sec:wire}.\\ 

We consider a pair of such coupled nanowires in the presence of a constant transverse bias $V_T$ and an out of plane magnetic field $B$ (as shown in Fig.~\ref{fig:circ}(a)). Current flow in the direction perpendicular to the wires occurs due to tunneling of electrons between wires. Reversing the sign of $V_T$ changes the direction of the current but preserves the underlying mechanisms. 
The out of plane magnetic field not only increases the energy levels of the electrons within each wire, it also increases the effective barrier height. On the other hand, application of a transverse voltage between the wires establishes a potential difference between them making the effective barrier between the wires skinnier and thus, more amenable for tunneling. 
We model the combined effects of $V_T$ and $B$ using the following interwire Hamiltonian.\\

\begin{align}
\begin{split}
  \mathcal{H}_{inter}  =&\frac{1}{2m^*}\Big(p_x-eB(y-\frac{p_x\omega_c}{m^*\Omega_B^2}) \Big)^2 +\\
  &\frac{p_y^2}{2m^*}-eV_T\frac{(y-y_{0})}{d-y_0}+E_0-\mu_BgBs
    \end{split}
    \label{eqn:hinter}
    \end{align}
where $d$ is the interwire spacing,  and $V_T$ is the transverse voltage applied between the wires.
The boundary of the first wire is at a distance $y_0$ from the center of the first wire. It marks the  beginning of the interwire potential.
The shift in guiding center of the electrons with the application of a magnetic field
is accounted for by $y\rightarrow y-{(p_x\omega_c)}/{(m^*\Omega_B^2)}$, and $E_0$ is the height of the potential barrier at $B=0$. 
The interwire Hamiltonian $\mathcal{H}_{inter}$ is discussed further in Section~\ref{sec:int} of the Appendix. 
We assume that the electron's energy and canonical momentum along the wire are conserved. We choose parameters such that the system is maintained in a regime of tunneling from a bound state in the first wire into the continuum of the next wire. In the reminder of this work, we set g = 0.62 and $m^*=1.9 m_e$, following Ref.
 \cite{annadi2018}.
\\


\begin{figure}[!hbtp]
    \centering
    \includegraphics[scale=0.73]{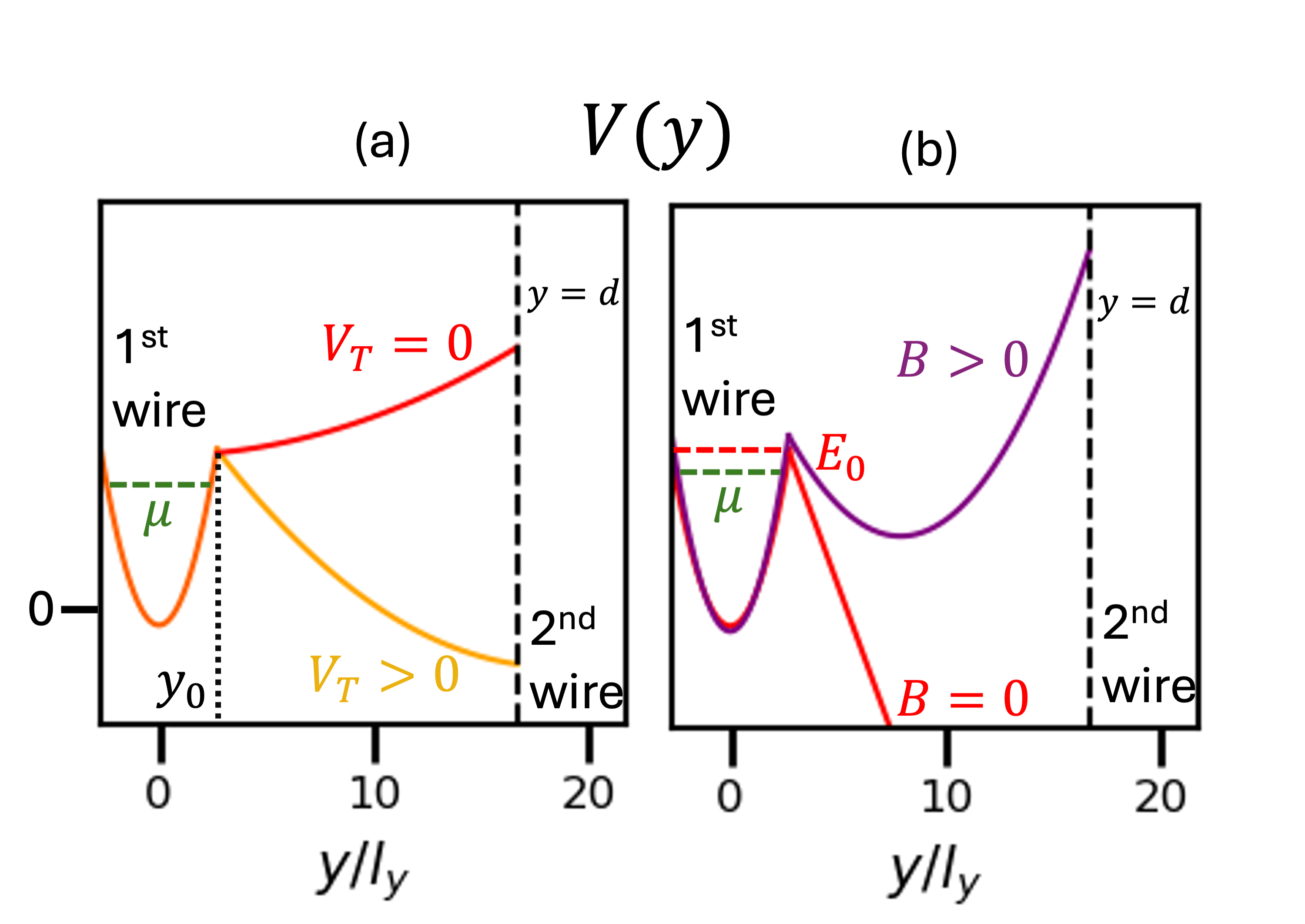}
    \caption[short]{The potential landscape as a function of position. (a) Change in the potential landscape with respect to 
    $V_T$ for constant $B$  and $k_x=0$. The red curve represents a potential with $V_T=0$ while the yellow curve represents a potential with $V_T>0$. $y_0$ which is the point that marks the boundary between the wire and the interwire potentials is shown. $(b)$ Change in the potential landscape with respect to $B$ for $k_x=0$, $s=\uparrow$ and $V_T>0$. The red dotted line depicts $E_0$ which is the height of the potential barrier at $B=0$}
    \label{fig:enter-label}
     \label{fig:ldscp}
\end{figure}

 \subsection{Calculation of Tunneling Rate}
 In order to calculate the tunneling rate, we turn to the Wentzel–Kramers–Brillouin (WKB) approximation. This technique has shown considerable success in predicting tunneling through potential barriers of various kinds~\cite{jelic2012,politzer1966,bayoumi2016,jelic2012,brinkman1970,manion1991}. 
We first estimate the transmission probability for a given initial state, followed by an integral over all of the occupied states in the first wire. 
 The tunneling rate $\Gamma$ is given by
 \begin{align}
 \Gamma(V_T,B,\epsilon)&=\Omega_B \mathcal{T}(\epsilon)\\
 &=\Omega_B e^{-2\int_{y_i}^{y_f}dy\sqrt{(2m^*/\hbar^2) (V(y)-\epsilon}) }
 \label{eqn:rate}
 \end{align}
where $\mathcal{T}(\epsilon)$ is the transmission probability through a single barrier. In the semi-classical picture, $\Omega_B$ is the frequency of electrons colliding against the walls of the wire. The effective potential $V(y)$ consists of two pieces, $V_{wire}$ and $V_{inter}$, both of which can be derived from
\begin{align}
V(y)=\mathcal{H}(y)-\Big(\frac{p_x^2}{2m^*}+\frac{p_y^2}{2m^*}+\frac{p_z^2}{2m^*_z} \Big) 
\end{align}

In the limit that $E_0 \rightarrow \infty$, $\mathcal{T}(\epsilon)$ is independent of applied fields. However, we are specifically interested in 
the case where the chemical potential is near the top of the potential barrier. 
While the increase in  electric field thins the barrier and increases tunneling current, the magnetic field plays a more complex role in the calculation. Increasing the magnetic field increases the tunneling attempt frequency of electrons within the wires, which sets the energy scale for the tunneling matrix element.  At the same time, the walls become more steep and the effective tunneling barrier for the electrons becomes larger (see Fig.~\ref{fig:ldscp}) which reduces the transmission probability. Finally, as shown in Fig.~\ref{fig:en}, magnetic depopulation leads to further expulsion of occupied states which in turn reduces the net tunneling current.

The quantum number  $k_x$  relates to the canonical x-momentum of a specific eigenstate. For our choice of gauge, $k_x$ is conserved. Turning on a magnetic field, 
the potential minimum shifts 
according to
Eq.~\ref{eqn:en}. Also, the barrier slope changes in a manner that is momentum dependent in such a way that it breaks $\pm k_x$ symmetry (see Eq.~\ref{eqn:hinter}).

Once the WKB integral is calculated for a single tunneling process, the total tunneling current is found by evaluating individual current contributions for each eigenstate and adding up contributions of all occupied energy states in the first wire. The tunneling conductance is then found by taking the derivative of the total current with respect to $V_T$.\\
\begin{align}
G_{\perp}=\frac{\partial}{\partial V_T}\Bigg(-e\int_{-\infty}^{\infty}d\epsilon \Omega_B\mathcal{T}(\epsilon)f_D(\epsilon)\frac{\partial n(\epsilon)}{\partial \epsilon}\Bigg)
\end{align}

For certain ranges of fields, a quasi-bound state appears in between the wires (see purple curve in panel (b) of Fig.~\ref{fig:ldscp})  where some energetic electrons are temporarily captured (see Ref~\cite{santiago2003}  for further discussions on this quasi-bound state). In such cases, the electrons must tunnel through two barriers to reach the next wire. Here,
we calculate the net conductance between the two wires as $G=G_1G_2/(G_1+G_2)$ where $G_1$ is the tunneling conductance between the first wire and  the potential minimum and $G_2$ is the tunneling conductance between the potential minimum and the second wire. Such processes, however, significantly diminish tunneling rate. Less energetic electronic states must tunnel through the entire interwire potential which also significantly suppresses their probability for tunneling. In contrast, states that obey  the following inequality (see yellow curve in panel (a) of Fig.~\ref{fig:ldscp})
\begin{equation}
E_{n_y}(B,k_x) \geq V_{inter}(V_T,B,y=d,k_x,s)
\label{ineq}
\end{equation}
experience only a single barrier which is made skinnier by $V_T$ and thus have a larger propensity to tunnel. It follows that the minumum voltage required for significant tunneling is given by\\  
\begin{eqnarray}
eV_T^*(k_x)\geq && \frac{1}{2}m^*\omega_c^2\Big(d-\frac{p_x\omega_c}{m^*\Omega_B^2}\Big)^2 - \hbar k_x\omega_c\Big(d-\frac{p_x\omega_c}{m^*\Omega_B^2}\Big) \nonumber \\ 
&&-E_{n_y}(B,k_x)+E_0-\mu_B gBs
\label{eqn:bdry}
\end{eqnarray}
where $\omega_c=eB/m^*$ is the cyclotron frequency and $E_{n_y}(B,k_x)$ is the energy of the state occupied by the electron within the wire as defined in Eq.~\ref{eqn:en}. Eq.~\ref{eqn:bdry} gives the critical voltage $V_T^*$ at which states with canonical x-momentum $\hbar k_x$ first experience significant tunneling. We calculate the number of states that are above $V_T^*$  ( and hence satisfy Eq. \ref{ineq}) using\\
\begin{align}
\begin{split}
     N(V_T,B)=&\sum_{n_y,s}\int_{-k_F}^{k_F}dk_x\Theta(\mu-E_{n_y}(B,k_x))\\&\times\Theta(E_{n_y}(B,k_x)-\mu+V_{inter}(V_T,B,y=d,k_x))\\
     &\times\frac{\partial n}{\partial k_x} f_D(E_{n_y}(B,k_x),\mu)
     \label{eqn:N}
\end{split} 
\end{align}
From Eq.\ref{eqn:N}, we can calculate the density of states available for tunneling. This allows us to identify the number of states that become accessible for tunneling at specific values of $V_T$ and $B$. This in turn facilitates a deeper understanding of the underlying mechanisms governing the field-dependence of transverse conductance.  



While the two-wire system cannot support a true metal-insulator phase transition, an array of wires can.   
Because the metal-insulator transition is characterized by a scalar order parameter with short-range interactions, leading to an Ising character to its phase transitions,\cite{liu2016,jayaraman1970,kadanoff1967,castellani1979,lu2021,limelette2003}  a phase transition is forbidden in a single wire, or even the two-wire system calculated here.  
However, as more wires are added, the system evolves into an effective 2D metamaterial, and the metal-insulator crossover can become a true phase transition as in the Ising model.
Strong evidence for this metal-insulator phase transition has been found in nanowire arrays sketched on LAO/STO, in which the conductance changes by up to four orders of magnitude \cite{ranjani2024}.
As with any first order phase transition, we expect to observe hysteresis when driving the nanowire array back and forth between the conducting and insulating phases.

\section{Results}
\begin{figure*}[]
 \centering
\includegraphics[scale=0.65]{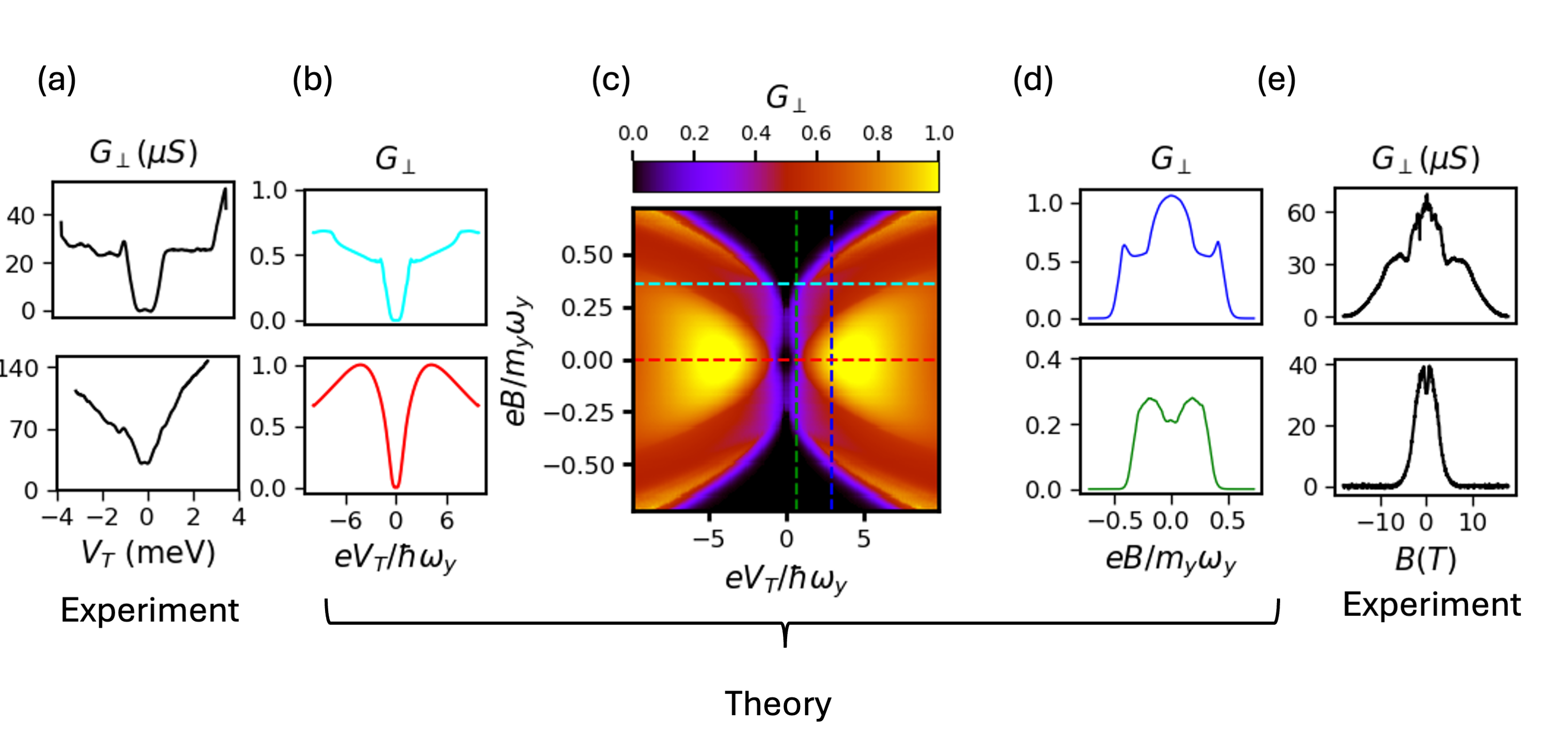}
\caption{(a) Transverse conductance plotted as a function of $V_T$ obtained from experiment. The lower plot corresponds to the $B=0$ Tesla linecut. (b) Transverse conductance from theory plotted as a function of $V_T$ in dimensionless units. (c) Transverse conductance plotted as a function of magnetic field and potential difference in dimensionless units.  $G_{\perp}$ is rescaled with respect to to the maximum value in the conductance map and is also dimensionless. (d) Transverse conductance from theory plotted as a function of $B$ in dimensionless units. (e) Transverse conductance plotted as a function of $B$ obtained from experiment. The lower plot corresponds to the $V_T=0$ linecut. In the theoretical panels (b), (c) and (d), the parameters chosen are $\mu/\hbar \omega_y=1.28$, $E_0/\hbar \omega_y=1.58$, $d/l_y=7.06$.}
    \label{fig:cmap}
\end{figure*} 

 \begin{figure*}[]
\includegraphics[scale=0.7]{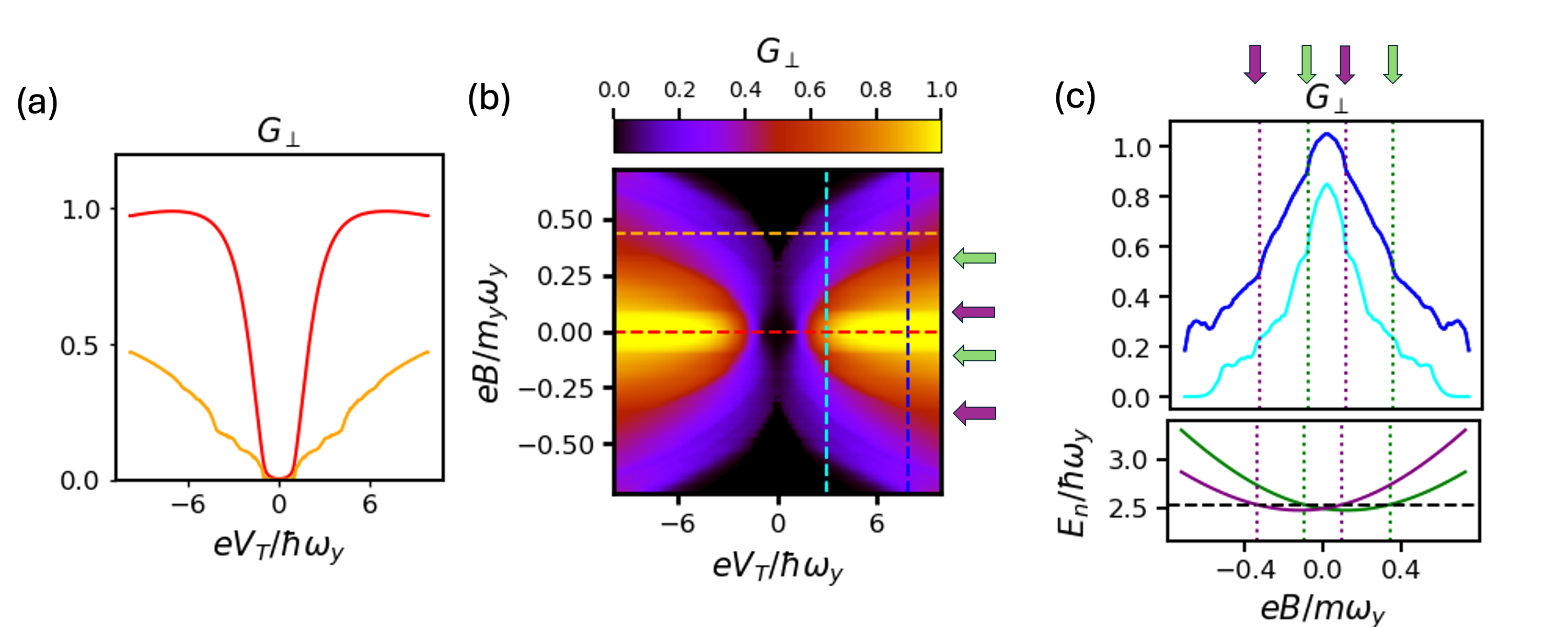}
\caption{Transverse conductance plots for wires with $n_y=0,1,2$ occupied energy levels. (a) Transverse conductance in dimensionless units plotted as a function of $V_T$ for the magnetic fields denoted by horizontal dashes in panel (b).  (b) Transverse conductance map plotted as a function of $V_T$ and $B$ in dimensionless units. (c) Transverse conductance in dimensionless units plotted as a function of $B$ for the values of $V_T$ denoted by vertical dashed lines in panel (b).  The corresponding energy levels in the wire for $n_y=2$ are also shown as a function of $B$.  In all the above plots, $G_{\perp}$ is rescaled with respect to to the maximum value in the conductance map and is also dimensionless. The parameters chosen are $\mu/\hbar \omega_y=2.54$, $E_0/\hbar \omega_y=2.95$ and $d/l_y=7.06$. }
    \label{fig:mgdp}
\end{figure*} 

\begin{figure*}[]
\includegraphics[scale=0.7]{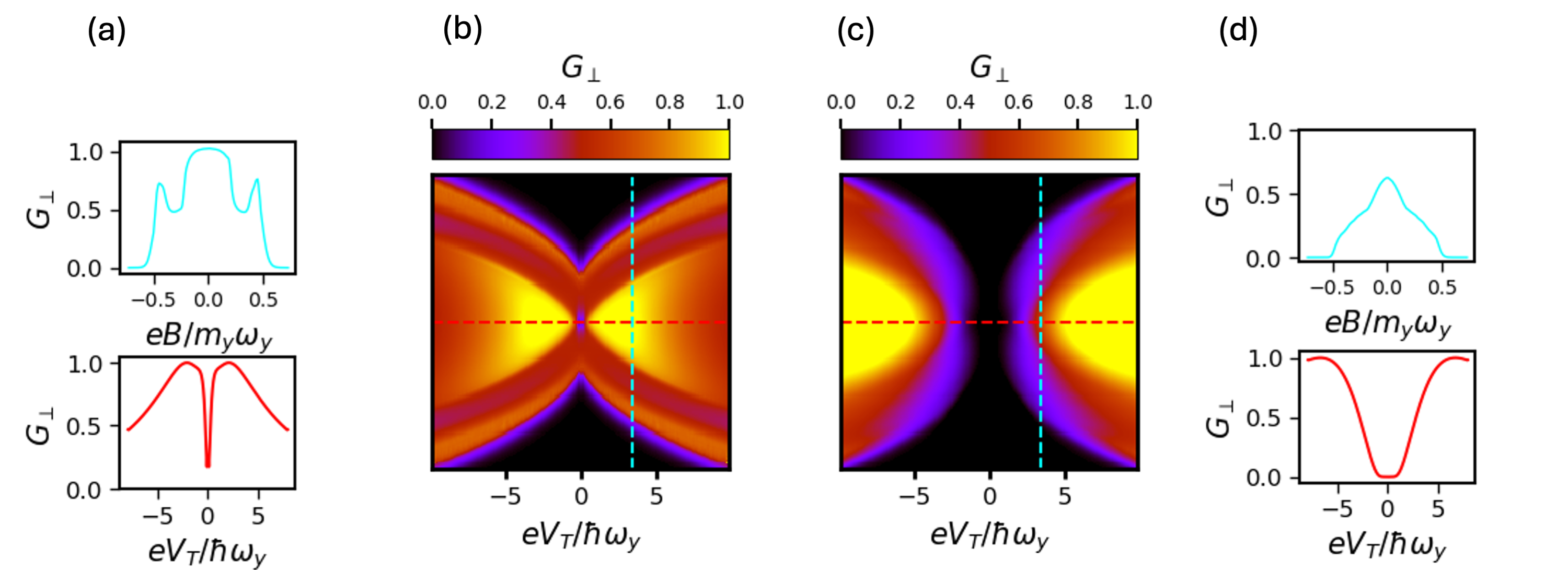}
\caption{Here, we compare the conductance maps between wires with different potential barriers $E_0$. Panels (a) and (b) represent a pair of wires with $E_0/\hbar \omega_y=1.38$ whereas panels (c) and (d) represent a wire with $E_0=1.77$. $G_{\perp}$ is rescaled with respect to the maximum value in the conductance map and is also dimensionless. The parameters chosen are $\mu/\hbar \omega_y=1.18$, $d/l_y=7.06$.}
    \label{fig:height}
\end{figure*}

\begin{figure}[htb]
    \centering
    \includegraphics[scale=0.9]{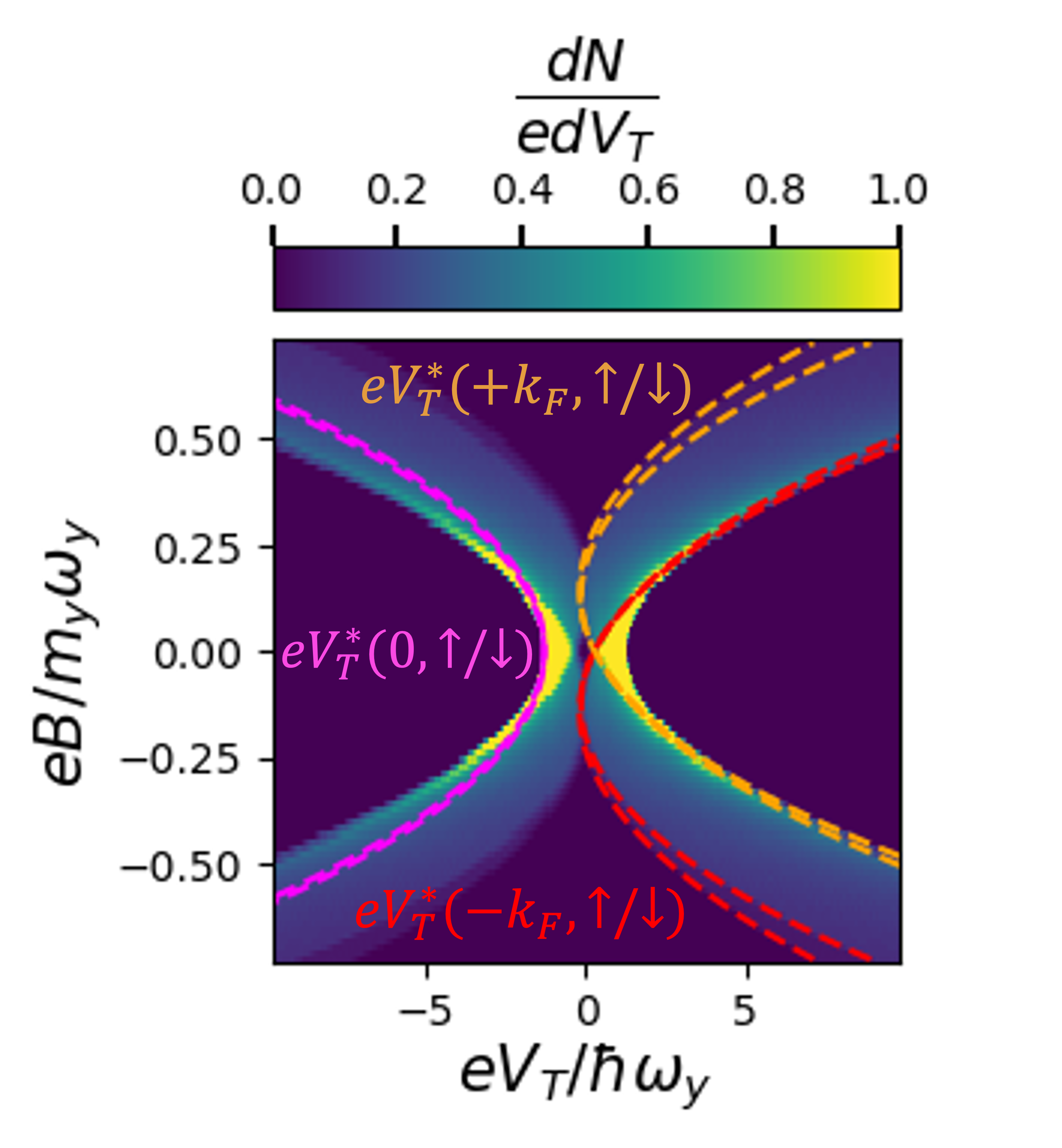}
    \caption{$(1/e)dN/dV_T$ is the density of states available for tunneling at a given $V_T,B$ normalized with respect to its maximum value. The parameters chosen are $\mu/\hbar \omega_y=1.28$, $E_0/\hbar \omega_y=1.58$, $d/l_y=7.06$. }
    \label{fig:num}
\end{figure}

Significant tunneling is only possible for a given state if it obeys the inequality in Eq.~\ref{eqn:bdry}. Using this equation, we find the distribution of states available for tunneling in a small neighbourhood around $eV_T^*(k_x)$ as a function of $B$ as shown in Fig.~\ref{fig:num}. Here, ~$N(V_T,B)$ refers to the number of states that are capable of tunneling as described in Eq. \ref{eqn:N}. $(1/e)dN/dV_T$ denotes the density of states that cross the threshold for tunneling at $V_T,B$. The orange and red dashed lines represent the critical potential difference for the Fermi momenta $k_x=\pm k_F$, $s=\uparrow,\downarrow$ as in Eq.\ref{eqn:bdry}. For $B>0$, it is seen that $eV_T^*(+k_F,\downarrow)$ is a lower bound for the minimum voltage needed for tunneling. Also, the $eV_T^*(-k_F)$ curves roughly demarcate the region where the density of states drops to zero. The same description holds for B<0 but with the momentum and spin signs reversed. At $B=0$, we see that the $eV_T^*(\pm k_F,s=\uparrow/\downarrow)$ curves meet at the point that corresponds to the smallest voltage needed for tunneling. The largest electric field is required by the least energetic state, namely $k_x=0$ (denoted by the pink curve in Fig.~\ref{fig:num}). For small $V_T$, none of the states are able to tunnel through as the barrier is too large and the density of states for tunneling is zero. For large $V_T$, all occupied states already participate in tunneling and here too, the density of states activated for tunneling goes to zero. 

The transverse conductance is plotted as a function of electric and magnetic fields in Fig.~\ref{fig:cmap} along with corresponding linecuts. The linecuts predicted by the model (panels (b) and (d)) are compared with those from experiments (panels (a) and (e)) on an array of nanowires sketched on the LAO/STO interface. 
Panel (b) shows the dependence of $G_{\perp}$ on $V_T$, 
taken through the horizontal dotted lines in panel (c) of Fig.~\ref{fig:cmap}.
Increasing $V_T$ initially increases conductance as more electronic states cross the threshold for tunneling. This drives an insulator-to-metal crossover (or a transition for an array of wires). 
For $B=0$ (the bottom graph in panel (b)),
the conductance reaches a maximum 
at $k_x=0$ (shown by the pink dotted line in Fig.~\ref{fig:num}) 
beyond which no new states are available for tunneling and the conductance decays. For sufficiently large $B>0$, 
(the top graph in panel (b) of Fig.~\ref{fig:cmap}), 
significant tunneling starts beyond $eV_T>eV_T^*(k_F)$
(corresponding to the orange dotted line in Fig.~\ref{fig:num}). 
The newly activated density of states attains a maximum when $eV_T \sim eV_T^*(-k_F)$ 
(the red dotted line in Fig.~\ref{fig:num}).

The vertical linecuts in panel (d) of Fig.~\ref{fig:cmap} show how $G_{\perp}$ depends on $B$ in theory and experiment. One sees that when $V_T \rightarrow 0$, there is a prominent minimum in $G_{\perp}$ at $B=0$ due to the fact that a linear momentum coupled $B$ term in the effective potential (as seen in Eq. \ref{eqn:hinter}) dominates for small $B$ and reduces the barrier thickness for electrons with $k_x>0$ for $B>0$ (and vice versa for $B<0$). This increases $G_{\perp}$ for increasing $|B|$. This is an important consequence of the momentum dependent term in the potential and is consistent with $G_{\perp}$ vs $B$ measurements in experiments seen in  Fig.~\ref{fig:cmap}(e).  At large $B$, the potential barrier rises as $B^2$ and dominates the physics regardless of the electron's momentum, making tunneling less likely.   The conductance reaches a maximum when crossing the region with the large density of activated states seen in Fig.~\ref{fig:num} and decreases when no new states remain to tunnel. When new spin channels are activated, $G_{\perp}$ attains a local maxima at these fields. This is seen in Fig. \ref{fig:height}(a). Finally, when the barrier is large enough for none of the states to tunnel, the conductance goes to zero.
 
In Fig \ref{fig:mgdp}, it is shown that when the chemical potential (i.e. backgate voltage) is adjusted such that multiple energy levels are occupied, it is possible to observe the effects of magnetic depopulation in the conductance plots. We see that a sharp drop in magnetoconductance occurs precisely when the bottom of a band crosses the chemical potential and becomes  unoccupied. These sharp features occur due to 
van-Hove singularities and are independent of $V_T$ as they do not depend on the structure of the interwire landscape. Instead, they are determined by the energy spectrum of the wires themselves. We also show in sec. \ref{sec:dos} of the Appendix that the inclusion of the $n_y=1$ subband results in a tangible increase in density of states activated for tunneling. This translates to an increase in conductance on the conductance map when states from the $n_y=1$ subband start to tunnel.

Finally, in Fig \ref{fig:height}, we examine the effects of increasing the potential barrier $E_0$ without changing the chemical potential. An insulating region opens up at the center of the conductance maps as electrons need a stronger bias $V_T$ to overcome the barrier. This is also seen in Fig. \ref{fig:dos2} in sec. \ref{sec:dos} of the Appendix where the tunneling density of states drops to zero at $V_T \rightarrow 0$ when the potential barrier is increased. Furthermore, we observe that the voltage at which the conductance attains a maximum is much larger when the potential barrier is higher. Thus, tuning the barrier height can give rise to either a decrease or enhancement of conductance. 

\section{Conclusion}
 We have considered a minimal model to describe tunneling between a pair of coupled nanowires that takes into account the effects of an electric and magnetic field-dependent potential barrier. 

We predict a magnetic field-dependent critical voltage above which significant tunneling of electrons begins. 

The application of an out of plane magnetic field to the coupled nanowires in combination with an applied transverse voltage
sculpts and changes the potential barrier so as to create 
characteristic peaks and valleys in the transverse differential
conductance maps.
One prominent consequence of this is the existence of a local minimum in conductance at zero field consistent with recent experiments on LAO/STO nanowire arrays \cite{ranjani2024}. 

The predictions of our model are also displayed for wires with different potential barriers and chemical potentials. We discuss the effects of multiple occupied subbands within the wire and demonstrate that a sharp change in conductance is expected at the magnetic field for which a subband is entirely depopulated. We also demonstrate the emergence of an insulating region in the conductance map for small electric fields when the potential barrier is increased. 

While a two wire system cannot support a true phase transition (being effectively one-dimensional), we argue that the metal-to-insulator crossover becomes a true phase transition in the limit of an array of wires.  

  The fact that we make minimal assumptions about the nature of interactions or material properties of the systems implies that the underlying mechanisms discussed are relevant to a larger class of coupled 1D systems. The results discussed here will be important to consider when one implements these structures in performing quantum simulation experiments. \\

\textbf{Acknowledgements} \par
SA acknowledges helpful conversations with Chris Greene, Jukka Varynen, Miguel Alarcón and Yuxin Sun. Acknowledgements
JL, PI, C-BE, and EC acknowledge support from from the Department of Energy under grant DOE-QIS (DOE
DE-SC0022277). EC acknowledges support from NSF Grant no. DMR-2006192. C.B.E. acknowledges support
for this research through the Gordon and Betty Moore
Foundation’s EPiQS Initiative, Grant GBMF9065 and a Vannevar Bush Faculty Fellowship (ONR N00014-20-
1-2844). Transport measurement at the University of
Wisconsin–Madison was supported by the US Depart-
ment of Energy (DOE), Office of Science, Office of Basic
Energy Sciences (BES), under award number DE-FG02-
06ER46327 (C.B.E.).

\appendix
\section{\label{sec:level2}The electron waveguide model of nanowires} \label{sec:wire}
First, we model the transport properties of a single nanowire with a voltage bias applied along the wire and a magnetic field applied perpendicular to the wire.  Just as in electron waveguides, the movement of electrons through the wire is ballistic. We treat the direction along a given wire ($\hat{x}$) as translationally invariant while the transverse ($\hat{y}$ and $\hat{z}$) directions posses convex potentials that can be approximated as parabolic wells. The simple harmonic approximation for the potential written in the Landau gauge is specifically chosen as it offers analytically solvable solutions to Schrodinger's equation. Applying a magnetic field along the $\hat{z}$ direction allows us to adjust the steepness of the parabolic wells while also breaking time translational invariance. As seen in Ref.~\onlinecite{annadi2018}, the Hamiltonian of a single wire is
\begin{align}
\mathcal{H}_{wire}=&\frac{1}{2m^*}(p_x-eBy)^2+\frac{p_y^2}{2m^*} +\frac{1}{2}m^* \omega_y^2y^2   -g\mu_BBs \nonumber\\&+\frac{p_z^2}{2m^*_z}+\frac{1}{2}m^*\omega_z^2z^2\\ \label{eqn:hwire1}
    =& \frac{p_x^2}{2m^*}\Big(\frac{\omega_y^2}{\Omega_B^2 }\Big)+\frac{p_y^2}{2m^*}+\frac{1}{2}m^* \Omega_B^2\Big( y-\frac{p_x \omega_c}{m^* \Omega_B^2}\Big)^2\nonumber\\& -g\mu_BBs+\frac{p_z^2}{2m^*_z}+\frac{1}{2}m^*_z\omega_z^2z^2
\end{align}
where $p_x$  is the canonical momentum operator along the $x$ axis, $p_y$ is the canonical momentum operator along the $y$ axis, g is the 
Landè g-factor, and $s$ is the spin of the electron. The angular frequency of the harmonic potential is $\omega_y=\hbar/m^*l_y^2$, where $l_y$ characterizes the transverse width of the wire. The effective frequency $\Omega_B$ is given by $\Omega_B=\sqrt{(\hbar/m^*l_y^2)^2+(eB/m^*)^2}$. Completing the square gives rise to a shift in the minimum of the wire's potential that couples with the magnetic field and the momentum of the wire. The extent of the shift is given by $(p_x\omega_c/m^*\Omega_B^2)$. The momentum dependent term can be thought of as altering the effective mass of the electron moving along the x direction. $m^{eff}_x(B)\rightarrow m^*_x(\Omega_B^2/\omega_y^2)$. The corresponding eigenvalues are\\
\begin{align}
\begin{split}
E_n(B)&=(n_y+1/2)\hbar \Omega_B+((2n_z+1)+1/2)\hbar\omega_z  \\& \quad -
g\mu_B Bs+\frac{\hbar^2k_x^2\omega_y^2}{2m^*\Omega_B^2}   
\end{split}
\end{align}
and the corresponding wave functions are given by \\
\begin{align}
\Phi_{n_y,k_x,s}(x,y,z)
=& N_{n_y,k_x}e^{-ik_xx-(m^*\Omega/2\hbar)(y-(\hbar \omega_c k_x/m^* \Omega_B^2))^2}\nonumber \\
& \times H_{n_y}\Bigg( \sqrt{\frac{m^* \Omega}{\hbar}}\Bigg(y-\frac{\hbar \omega_ck_x}{m^*\Omega_B^2}\Bigg)\Bigg)\nonumber\\&\times e^{-(m^*_z\omega_z/2\hbar)z^2}H_{2n_z+1}\Bigg(\sqrt{\frac{m^*_z\omega_z}{\hbar}z} \Bigg)
\end{align}

\begin{figure}
    \centering
    \includegraphics[scale=0.82]{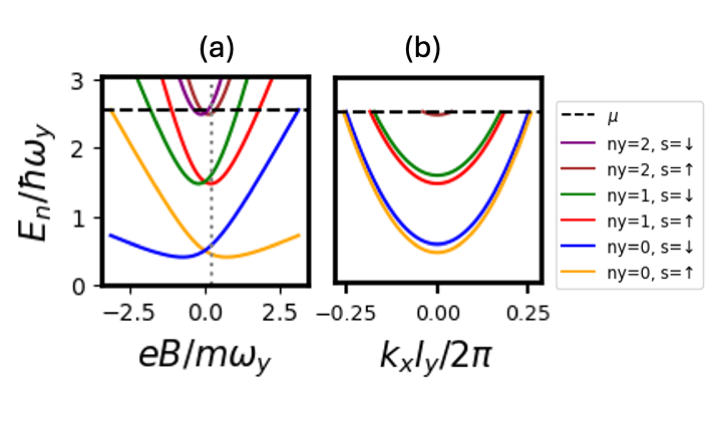}
    \caption{(a) depicts energy levels in the model as a function of $B$ for $k_x=0$. (b) depicts energy levels plotted as a function of $B$ for $eB/m\omega_y=0.2$}
    \label{fig:en}
\end{figure}

 It has been shown in~\cite{annadi2018} that transconductance measurements along the wire are accounted for by the energy levels predicted by Eq.~\ref{eqn:en}. Landauer's formula~\cite{datta2005} is an excellent description of parallel conductance seen in ballistic dissipationless nanowires. Four terminal measurements have confirmed that the resistance arises from the leads and not from the wire itself~\cite{depicciotto2001}. Every energy band contributing to the current imparts a conductance of $e^2/h$. This result is robust even in the presence of electron-electron interactions~\cite{maslov1995a}. \\
\begin{align}
 G_{||}=\frac{e^2}{h}\sum_{n_y,s} \bar{T}F_{T}(\mu-E_n(n_y,s)) 
\end{align}
where $\bar{t}$ encompasses any tunneling resonances or backscattering processes. Here, we set $\bar{t}=1$ for ballistic dissipationless transport within the nanowires. $F_T(E)$ is the thermal broadening function of the Fermi distribution \cite{datta2005} given by 
\begin{align}
    F_T(\epsilon,\mu)&=\frac{\partial}{\partial \mu}\Bigg(\frac{1}{e^{(\epsilon-\mu)/k_B T}+1}\Bigg).
\end{align}

 We can use this expression along with our Hamiltonian for the nanowire to calculate $G_{||}$ vs $B$ in isolated nanowires. We use Landauer's formula for calculating conductance of isolated nanowires which is given by: \\ \\

\begin{figure}[]
    \centering    \includegraphics[scale=0.7]{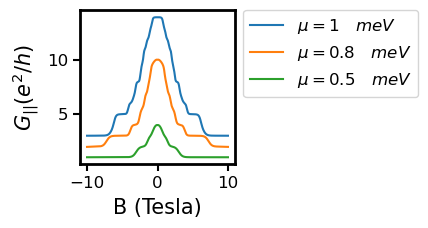}
    \caption{$G_{||}$ vs $B$ is plotted for different $\mu$.The calculation here uses $T=50$mK, The wire width used is $l_y=17$nm.}
    \label{fig:Gpl}
\end{figure}
The step-like drops in $G_{||}$ seen in Fig.~\ref{fig:Gpl} occur due to depopulation of the wire that occur when an increasing magnetic field pushes energy levels (as seen in Fig.~\ref{fig:en}) above the chemical potential thus depopulating them. The steps are smoothed out at higher temperatures due to thermal broadening. The overall form of the plot is consistent with transport measurements performed for nanowires sketched on the LAO/STO interface ~\cite{cheng2018}. These measurements demonstrate that magnetic depopulation of bound states within the wire account for the drop in conductance seen with increasing magnetic fields. Previous work~\cite{annadi2018} also demonstrates the success of using Eq.~\ref{eqn:hwire} to describe transconductance features in nanowires.  

\section{Interwire Hamiltonian} \label{sec:int}

From Eq.~\ref{eqn:hwire1}, we see that guiding centers of the electrons with non-zero momentum get shifted in the presence of a magnetic field. Ensuring a commensurate shift occurs in the interwire potential, the interwire Hamiltonian is then written as\\
\begin{align}
\begin{split}
  \mathcal{H}_{inter}  &=\frac{1}{2m^*}\Big(p_x-eB(y-\frac{p_x\omega_c}{m^*\Omega_B^2}) \Big)^2+\frac{p_y^2}{2m^*}+E_0\\ &\quad-\frac{eV_T}{d-y_0}(y-y_{0})-\mu_BgBs+\frac{p_z^2}{2m^*_z}+\frac{1}{2}m^*_z\omega_z^2z^2 \label{eqn:hinter1}
    \end{split}
    \end{align}

\begin{align}
\begin{split}
     y_0=&\frac{p_x\omega_c}{m^*\Omega_B^2}-\frac{p_x \omega_c}{m^*\omega_y^2}\\&+ \frac{1}{m^*\omega_y^2}\sqrt{p_x^2\omega_c^2+2m^*\omega_y^2\Big(E_0+\frac{p_x^2\omega_c^2}{2m^*\Omega_B^2} \Big)}
\end{split}
\label{eqn:y0}
\end{align}

Here, $d$ is the interwire spacing measured as the distance between the centers of two adjacent wires,  $V_T$ is the transverse voltage applied between the wires, $y_0$ is the location of the boundary between the wire and the interwire potential and is found solving for $y_0$ in $\mathcal{H}_{wire}(y_0)=\mathcal{H}_{inter}(y_0)$. 

We assume that only the lowest $z$ subband is occupied.
Within the wire, the magnetic field and the intrinsic curvature of the wire influence the energy eigenmodes. The corresponding Hamiltonian for the wire is given by $\mathcal{H}_{wire}$ in Eq.~\ref{eqn:hwire}. 
Because the wire is metallic, the potential drop across the wire in the transverse direction is negligible.
Beyond $y=y_0$, the potential landscape is determined by the choice of electric and magnetic fields applied. The larger the magnetic field, the steeper the barrier. The stronger the electric field, the skinnier the barrier. Furthermore, we see in Eq.~\ref{eqn:y0} that the wire boundary not only shifts for a finite $k_x$ and $B$, the effective width of the wire also increases. Since the minimum of the potential shifts for eigenstates with non-zero $k_x$ and $B$, the model is constructed such that the boundary of the wire also shifts with it in order to ensure that the wire does not spontaneously ionize. 
\vspace{-1em}
\section{Varying chemical potential} \label{sec:mu}
\begin{figure}[htb]
    \centering
     \includegraphics[scale=0.78]{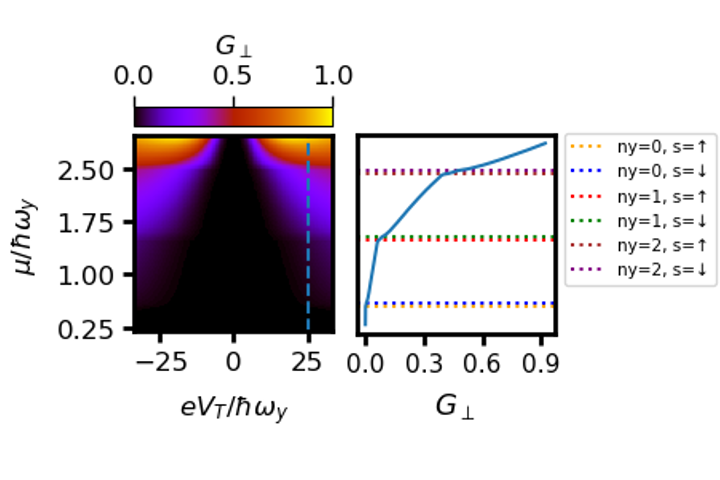}
    \caption{ Conductance map of $G_{\perp}$ with $eB/m\omega_y=0.1$ and $E_0/\hbar \omega_y=3.75$, $d/l_y=14.7$ for different $\mu$ and $V_T$. $G_{\perp}$ is rescaled with respect to to the maximum value in this map. The horizontal dotted lines denotes various van-Hove singularities that are present within the wire.  }
    \label{fig:bg}
\end{figure}
Here, the conductance map is plotted as a function of potential difference and chemical potential of the wires in Fig.~\ref{fig:bg}. It is seen that a triangular wedge shaped insulating region gives way to a conducting region upon increasing voltage or chemical potential $\mu$. increasing $\mu$ makes $\mu/E_0$ smaller and increasing $V_T$  makes the barrier skinnier. Both result in an increased current. This explains the shape of the insulating region in the conductance map. We also see bands across the conduction map at specific values of $\mu$. These bands occur precisely when $\mu$ crosses the bottom of an energy band. That is, $\mu=E_n(k_x=0,n_y=0,1,2,..)$. There exist van Hove singularities at these energies which significantly increase the tunneling and manifest as sharp changes in conductance. In general,  A prominent triangular insulating wedge is also seen in experiment which we demonstrate in \cite{ranjani2024}.
\section{Density of states activated for tunneling} 
\label{sec:dos}
\begin{figure}[hbtp!]
    \centering
\includegraphics[scale=0.55]{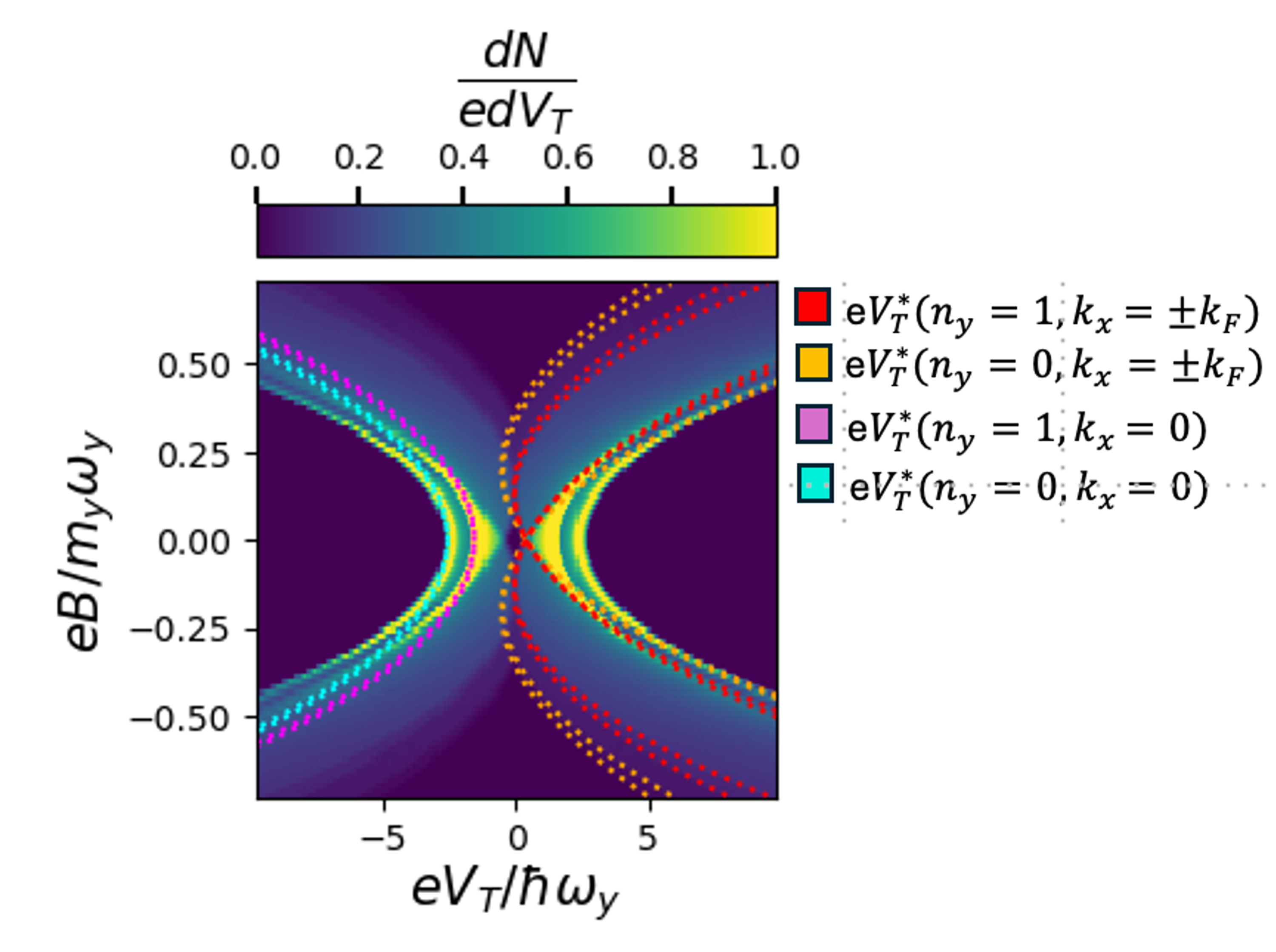}
    \caption{Map of density of states available for tunneling in wires with multiple occupied subbands with parameters $\mu/\hbar \omega_y=2.54$, $E_0/\hbar \omega_y=2.95$ and $d/l_y=7.06$}
    \label{fig:dosn=1}
\end{figure}

\begin{figure}[hbtp!] 
    \centering
\includegraphics[scale=0.45]{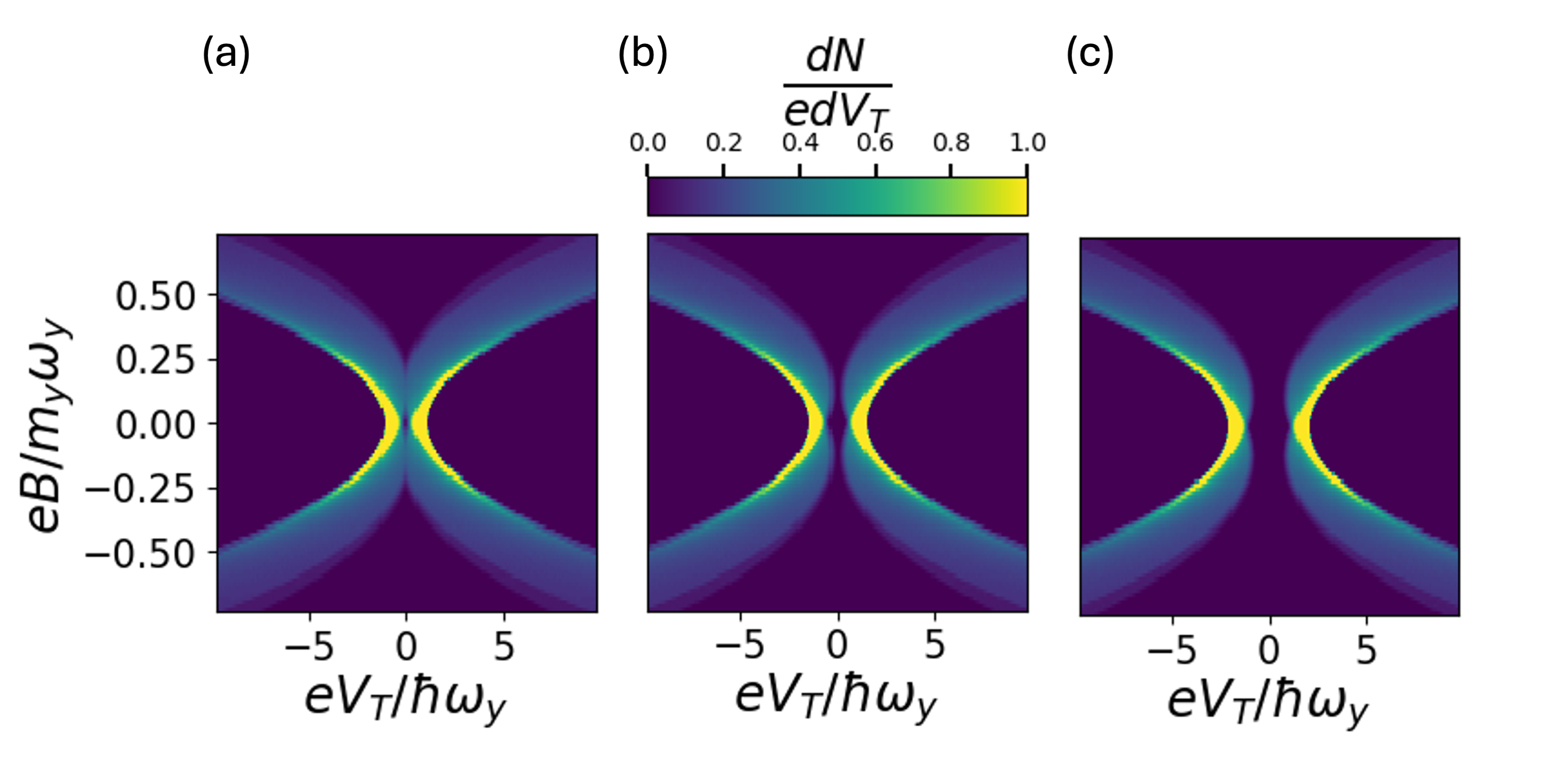}
    \caption{ Map of density of states activated for tunneling with (a) $E_0/\hbar \omega_y=1.38$, (b) $E_0/\hbar \omega_y=1.77$ and (c) $E_0/\hbar \omega_y=2.36$. $\frac{dN}{edV_T}$ is rescaled with respect to the maximum value in the conductance map and is also dimensionless. Other parameters chosen for all panels is $\mu/\hbar \omega_y=1.18$, $d/l_y=7.06$ }
    \label{fig:dos2}
\end{figure}

Here, we include maps for the density of tunneling states that correspond to the parameters defined in Fig[\ref{fig:mgdp},\ref{fig:height}].In Fig [\ref{fig:dosn=1}], when van Hove singularities ($k_x=0$ states) corresponding to $n_y=0,1$ (marked by cyan and pink colored dashed lines Eq.\ref{eqn:bdry} ) start to tunnel, the intensity of plots is larger at these fields. Also, when the $n_y=1, k_x=\pm k_F$ subband (marked in red dashed lines obtained from Eq.\ref{eqn:bdry} ) first participates in tunneling, the intensity of the density plots increases noticeably. 
In Fig [\ref{fig:dos2}], the emergence of an insulating region at $V_T \rightarrow 0$ is demonstrated for wires with a larger barrier height. This is expected as a larger potential barrier diminishes the tunneling rate and must be compensated by a larger electric field to enable tunneling. The intensity of the both maps drops to zero when no new states are present for tunneling despite increasing electric fields.  This explains why $G_{\perp}$ stagnates or decays for large electric fields as seen in panel (a) of Fig.[\ref{fig:mgdp}] and Fig.[\ref{fig:height}].

In Fig.[\ref{fig:dos2}], we see the effect of increasing the potential barrier between the wires. An insulating region appears around $V_T=0$ owing to the fact that a larger electric field is required to compensate for a larger potential barrier. This phenomenon is also observed in Fig. [\ref{fig:dos2}] where it is seen that the voltage at which the conductance attains a maximum is larger when $E_0$ is larger.
\clearpage
\bibliography{export2,EC-byHand}
\end{document}